\documentclass[fleqn,12pt,twoside]{article}
\usepackage{espcrc1}
\usepackage{graphicx}
\usepackage[figuresright]{rotating}

\newcommand{\AmS}{{\protect\the\textfont2
  A\kern-.1667em\lower.5ex\hbox{M}\kern-.125emS}}
\title{Baryon Resonance Dynamics in
$\pi N \to N V$ Reactions Near Threshold\thanks{Supported
by BMBF grant 06DR921 and GSI.}}

\author{B. K\"ampfer\address{Forschungszentrum Rossendorf,
                    PF 510119, 01314 Dresden, Germany},
        A.I. Titov\address{Advanced Science Research Center,
                  Japan Atomic Energy Research Institute,
                  Tokai, Ibaraki,  319-1195, Japan}\thanks{
                  On leave of absence from
                  Bogolyubov Laboratory of Theoretical Physics,
                  JINR Dubna, 141980, Russia.} and
        B.L. Reznik\address{Far Eastern National University,
                  Vladivostok 690000, Russia}
}

\begin{document}

\maketitle

\begin{abstract}
The $\rho - \omega$ interference in the exclusive reaction
$\pi N \to N' e^+ e^-$ is studied within a schematic model
including the established baryon resonances up to 1720 MeV.
Near threshold the interference can be used to separate
the isoscalar part of the electromagnetic current.
The role of various baryon resonances is highlighted.
\end{abstract}

\section{INTRODUCTION}

The {\bf H}igh {\bf A}cceptance {\bf D}i-{\bf E}lectron {\bf
S}pectrometer (HADES) \cite{HADES}, now becoming operational,
allows a new series of measurements being of high relevance for
contemporary strong interaction physics. Among the problems
addressed are heavy-ion experiments searching for further
indications of chiral symmetry restoration. HADES is well-equipped
to study also elementary hadron reactions since at GSI, besides
heavy-ion beams, proton and pion beams are available, and hydrogen
or deuterium targets can be used. This allows a detailed study of
the reactions $\pi^- p \to n e^+ e^-$ and $\pi^+ n \to p e^+ e^-$,
where a subtle $\rho - \omega$ interference is expected
\cite{Madeleine1,we,Madeleine2}. In these reactions the excitation
of intermediate baryon resonances is important. Thus, the baryon
resonance dynamics becomes accessible. We are going to show the
sensitivity of various observables on individual resonance
properties. A solid understanding of the excitations of the
nucleon, which constitutes the majority of luminous mass in the
universe, is still lacking as highlighted by the ''missing
resonances'' predicted by models but not yet identified. Also the
link of baryon resonance parameters to QCD lattice results
\cite{Leinweber} is still challenging.

\section{ISCALAR-ISOVECTOR INTERFERENCES IN
$\pi N\to N e^+e^-$ REACTIONS AS A PROBE OF BARYON RESONANCE DYNAMICS}

Coupled channel approaches \cite{Friman,Penner_Mosel} are well
suited for simultaneously analyzing the wealth of data on hadron
induced reactions on the nucleon. Since we restrict ourselves to a
narrow energy region around thresholds and to the special reaction
$\pi N \to N' e^+ e^-$ we resort to a schematic Born term model
with only $s$ and $u$ channels. The heart of the exclusive
reaction is the subprocess $\pi N \to N' V$, $V = \rho, \omega$
with subsequent decay $V \to e^+ e^-$. We take into account in the
isoscalar channel the nucleon and $P_{11}(1440)$, $D_{13}(1520)$,
$S_{11}(1535)$, $S_{11}(1650)$, $D_{15}(1675)$, $F_{15}(1680)$,
$D_{13}(1700)$, $P_{13}(1720)$, while the nucleon and
$P_{11}(1440)$, $D_{13}(1520)$, $S_{11}(1535)$, $S_{11}(1650)$,
$D_{15}(1675)$, $F_{15}(1680)$, $D_{13}(1700)$, $P_{13}(1720)$,
and the four $\Delta$ states $P_{33}(1232)$, $P_{33}(1600)$,
$S_{31}(1620)$, $D_{33}(1700)$ are accounted for in the isovector
channel. The interaction Lagrangians and parameters (cut-offs,
couplings, phases etc.) are listed in \cite{we}; the latter ones
are from \cite{Riska_Brown}.

If ones denotes the corresponding amplitude as
$T^{\pi^-p\to n e^+e^-} =
T^{\rm isoscalar} + T^{\rm isovector}$
the isospin rotated amplitude becomes
$T^{\pi^+n \to p e^+ e^-} =
T^{\rm isoscalar} - T^{\rm isovector}$ pointing to
different interferences in both reactions. Indeed, as exhibited
in Figure 1, below the threshold of
\begin{figure}[hb]
~\vskip -6mm
\centerline{
\includegraphics[width=15pc]{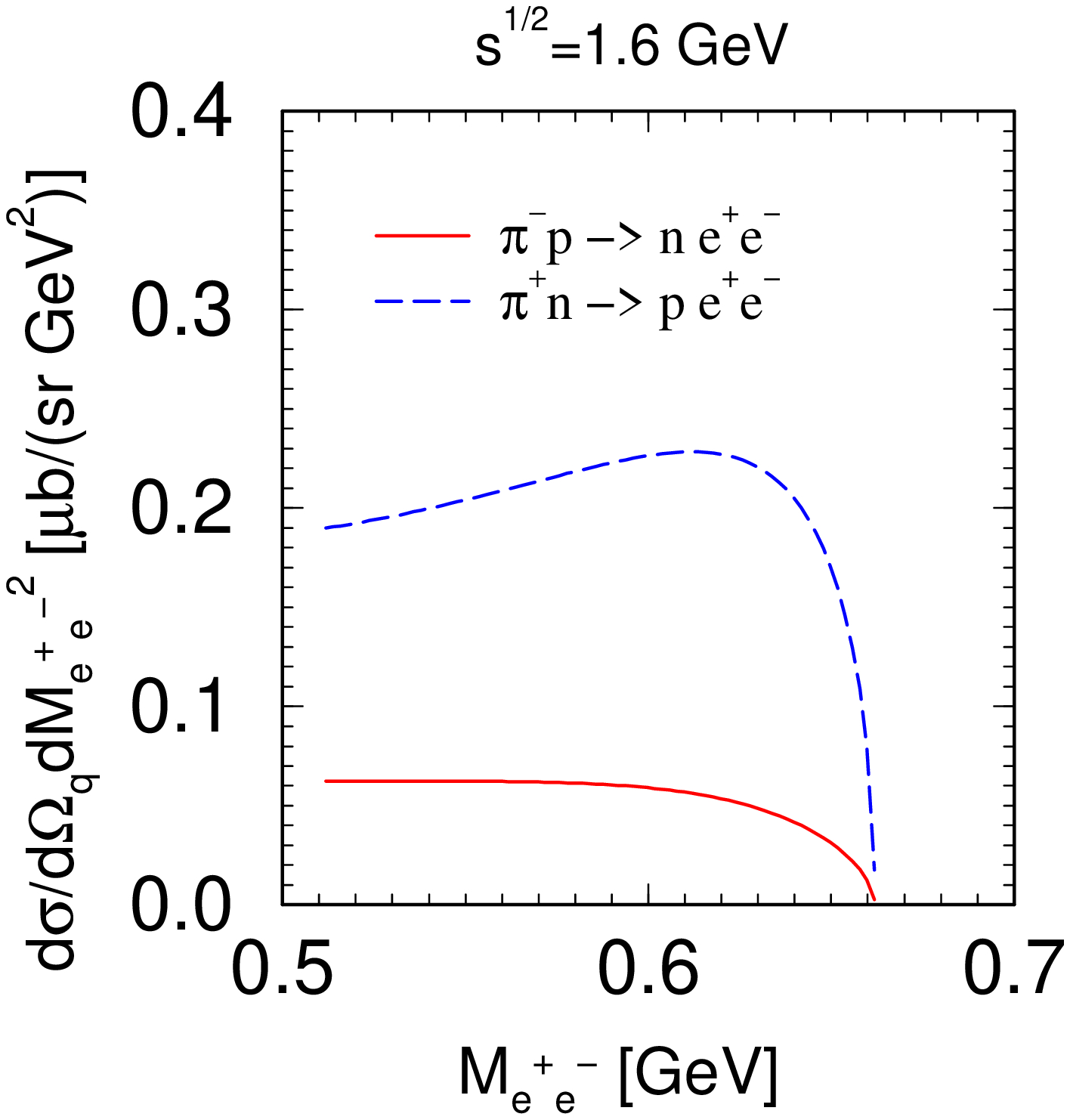} \hspace*{9mm}
\includegraphics[width=16pc]{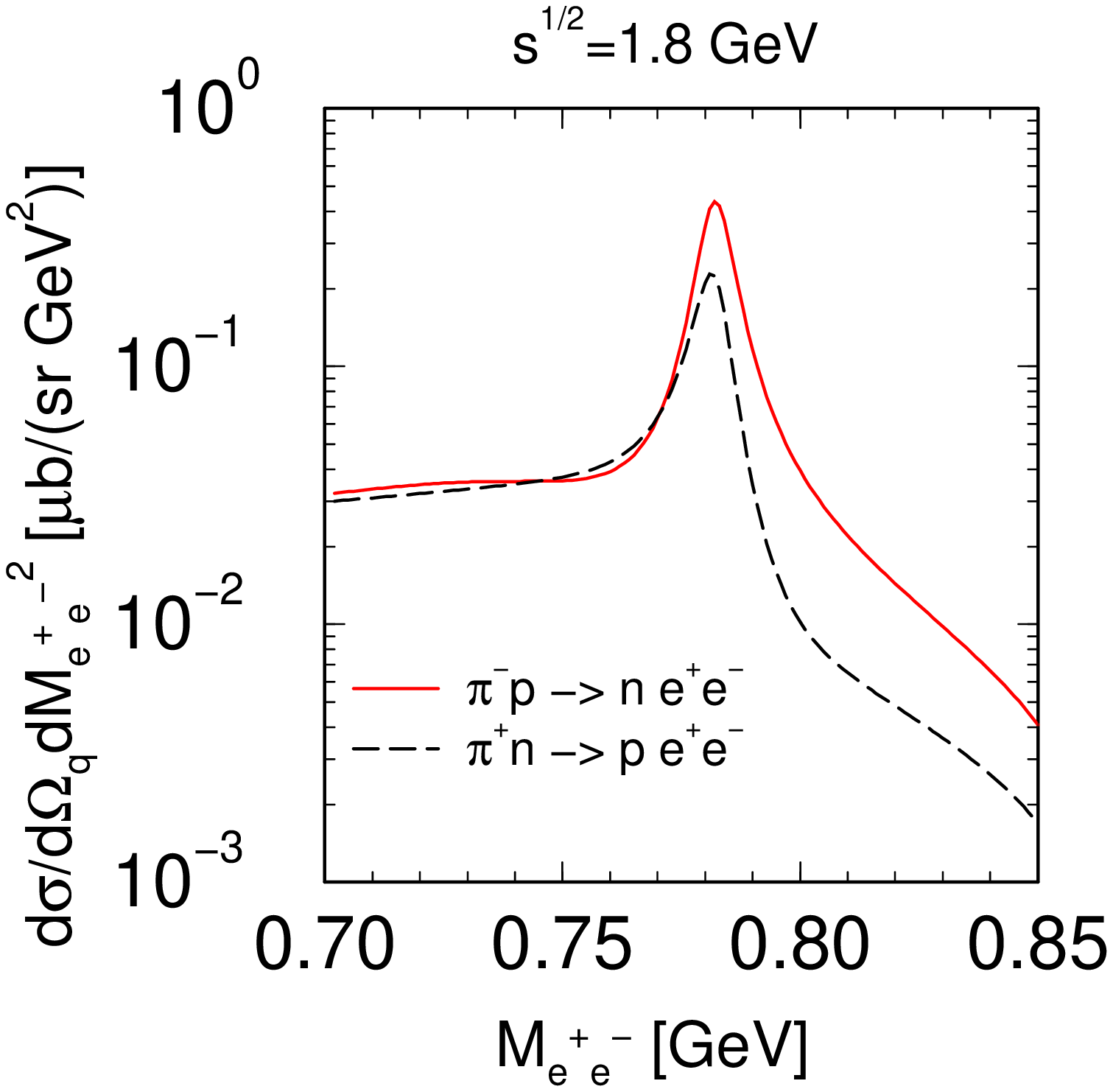}}
~\vskip 1mm
\centerline{
\includegraphics[width=15pc]{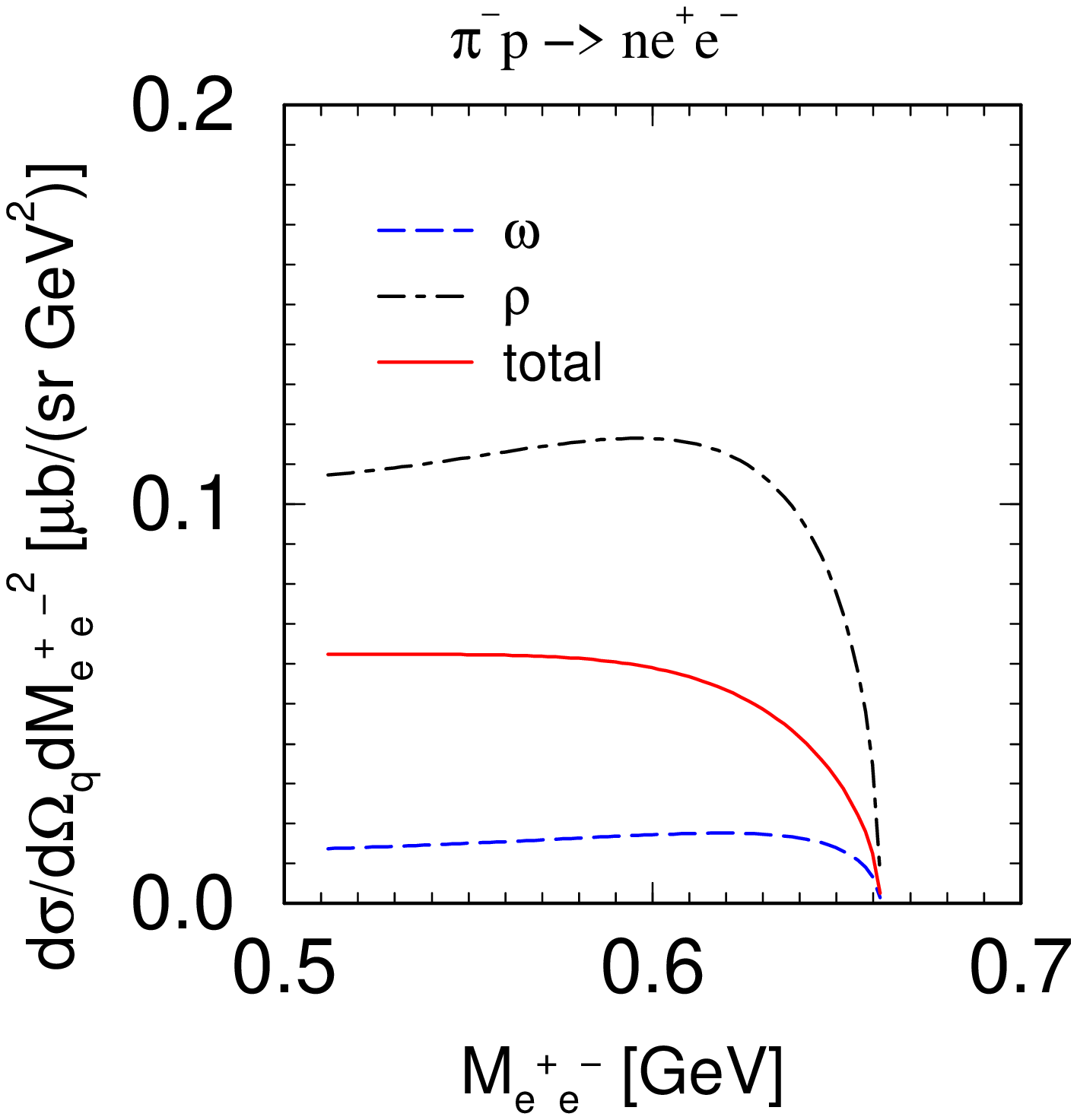} \hspace*{9mm}
\includegraphics[width=16pc]{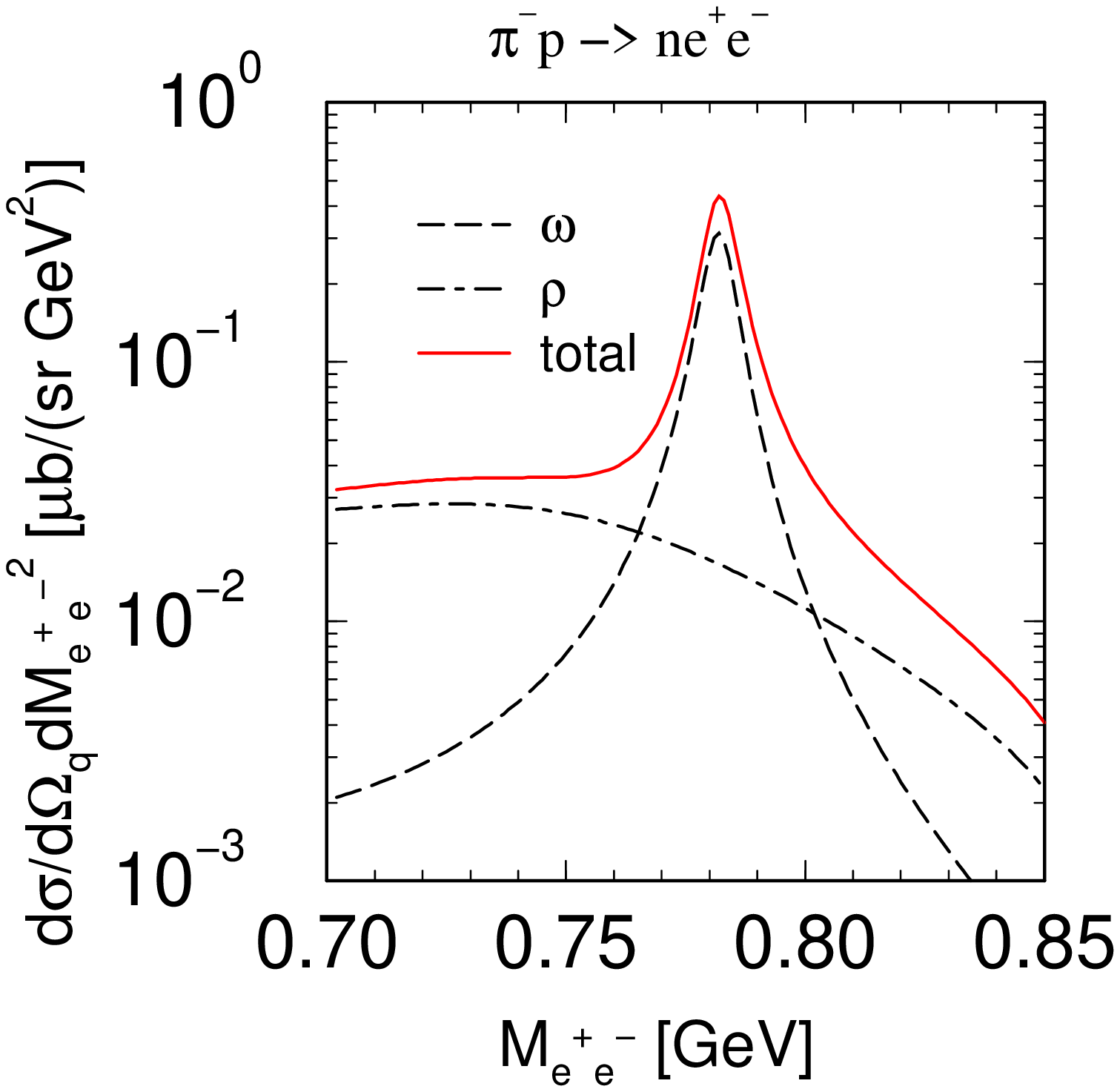}}
~\vskip -12mm
\caption{Differential cross sections below (left panels) and above
threshold (right panels) as a function of the invariant mass.
The lower row depicts separately the $\rho$ and $\omega$ channels.
The polar angle is $\theta = 30^o$ in the $\pi N$ center-of-mass
system.}
\label{fig2}
\end{figure}
$\sqrt{s} = 1.72$ GeV
the differential cross sections of both reactions are quite
different, while above threshold they are similar.
Below threshold, the destructive interference between the
$\rho$ and $\omega$ contributions in the reaction
$\pi^-p\to n e^+e^-$ are responsible for this behavior
(see Figure 2). The energy dependence of the differential
cross sections are displayed in Figure 3 at fixed invariant
mass of $M = 0.6$ GeV. With changing energy also the individual
contributions of the baryon resonance change, as shown in
Figure 4. We emphasize that the influence of the baryon
resonances changes with the angle of the outgoing
$e^+ e^-$ pair.

\begin{figure}[htb]
~\vskip -6mm
\centerline{
\includegraphics[width=15pc]{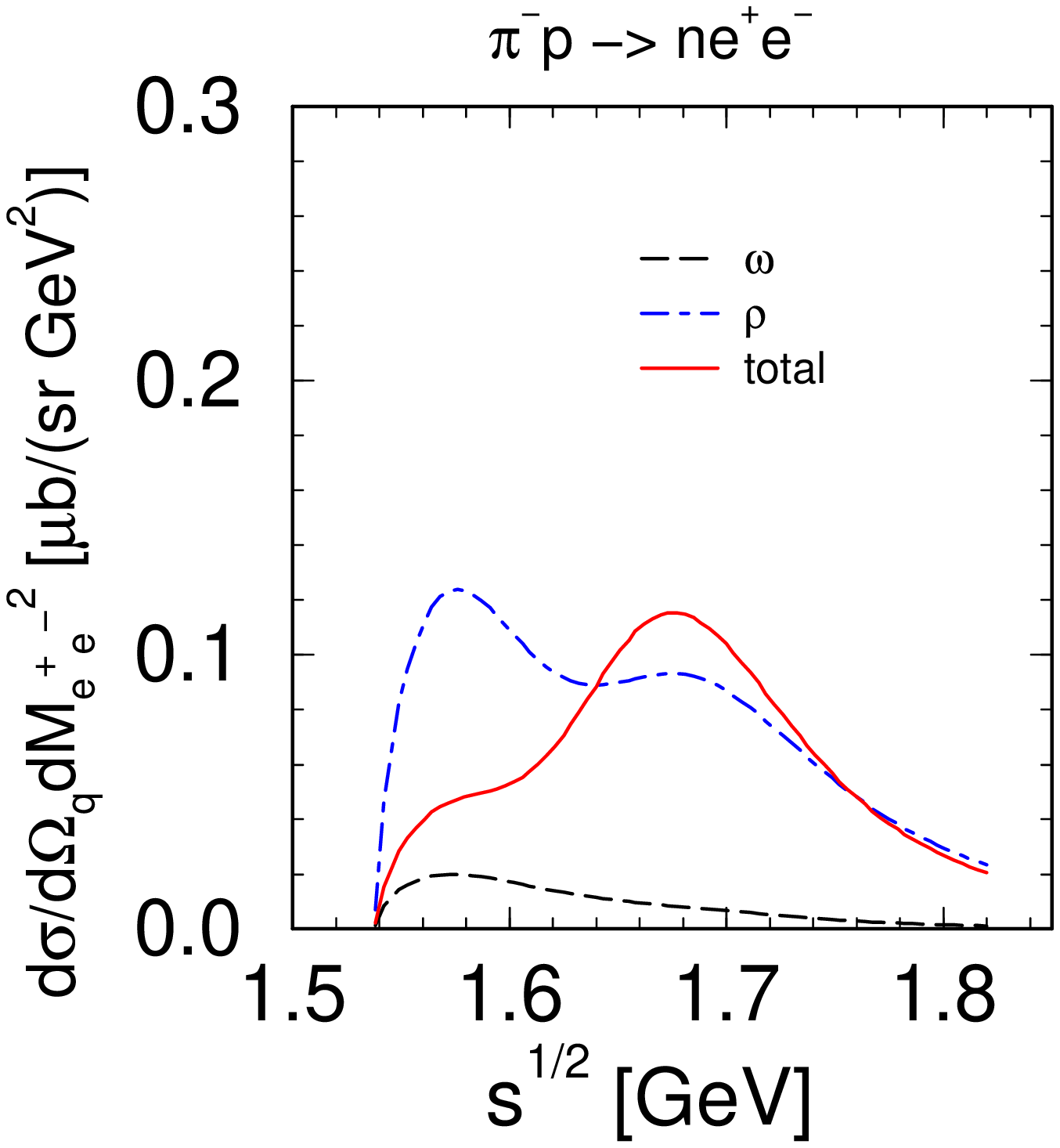} \hspace*{12mm}
\includegraphics[width=15pc]{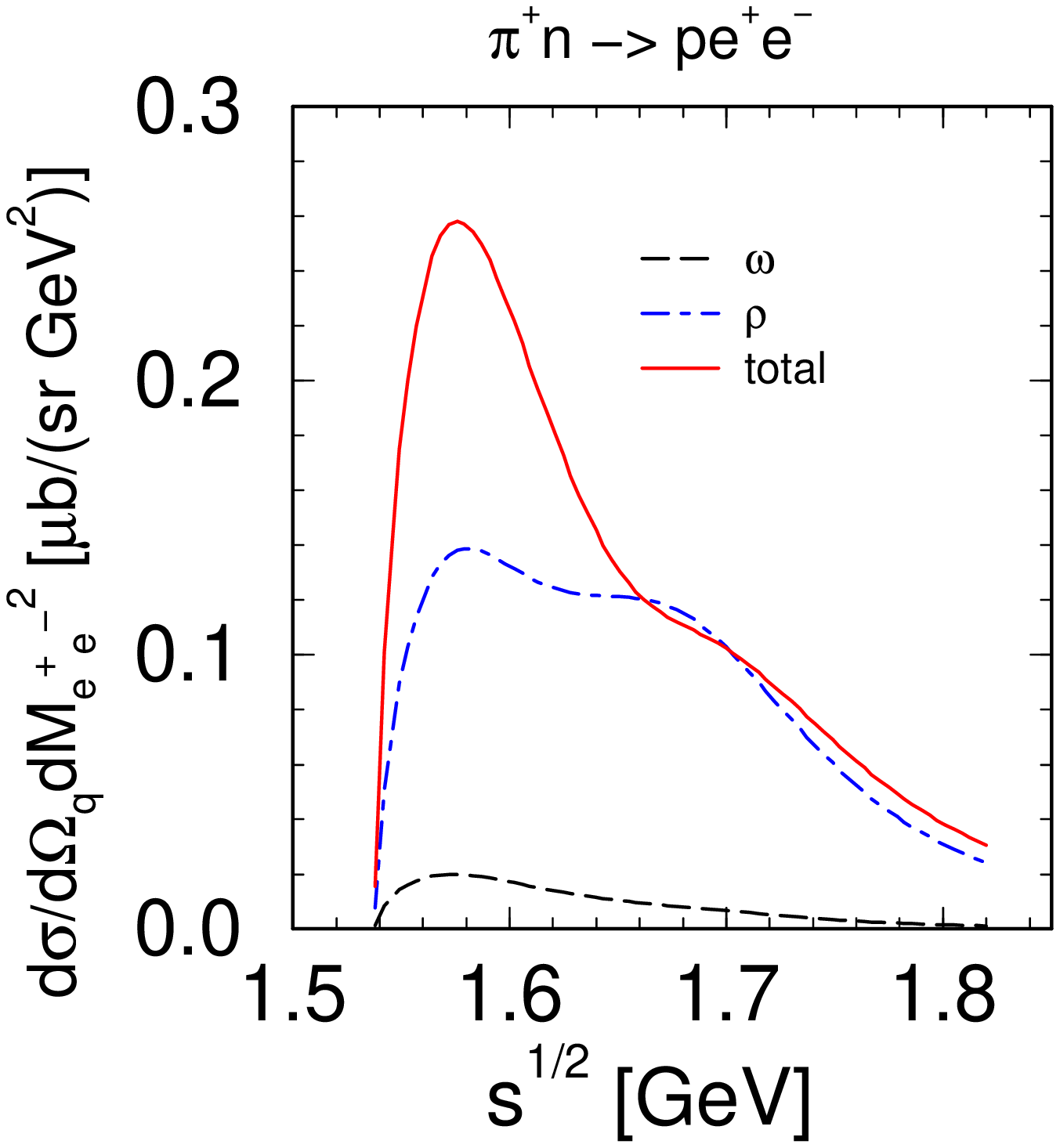}}
~\vskip -12mm
\caption{Energy dependence of differential cross section
at fixed invariant mass ($\theta = 30^o$).}
\label{fig3}
~\vskip 3mm
\centerline{
\includegraphics[width=17pc]{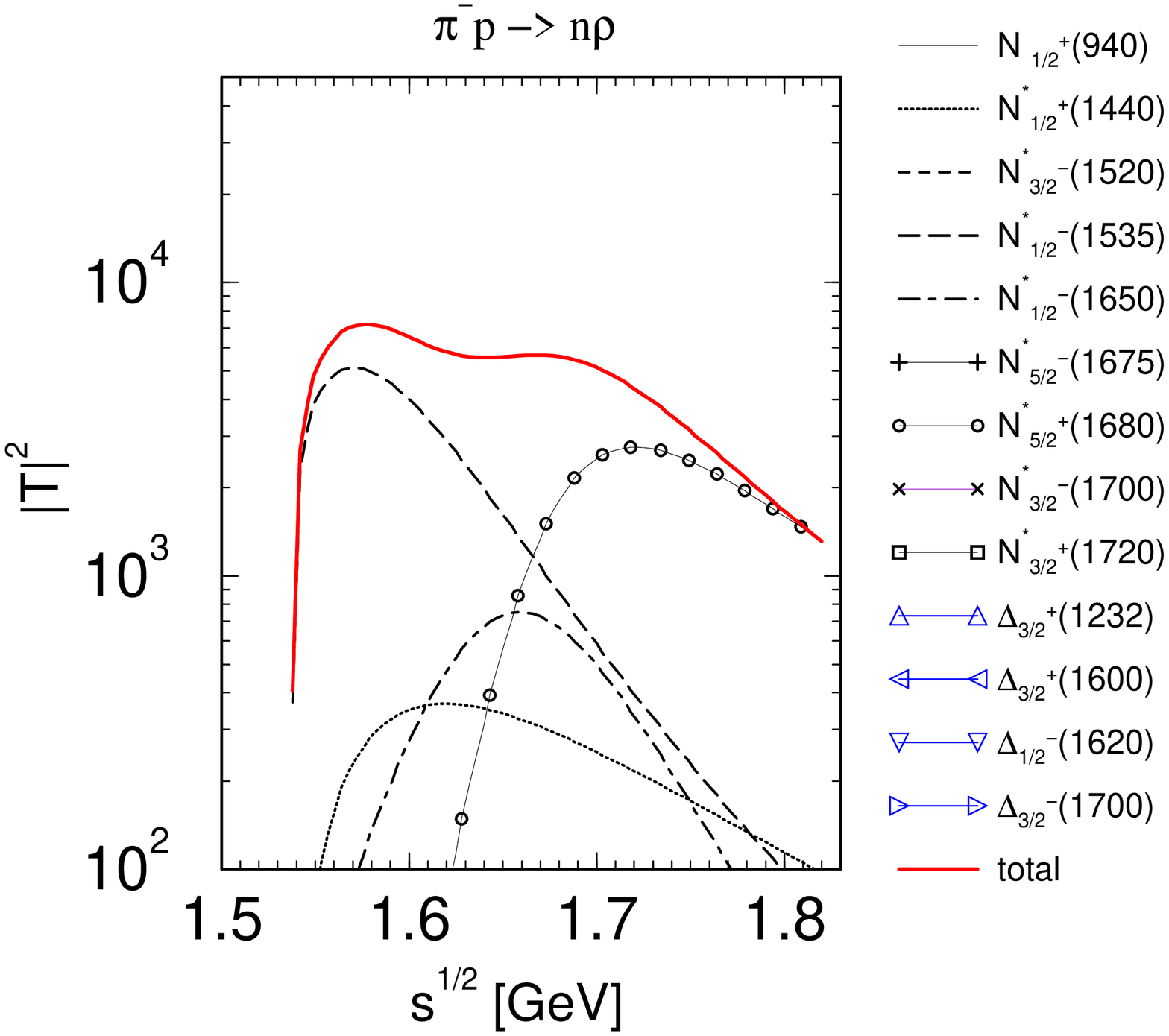} \hspace*{12mm}
\includegraphics[width=13.3pc]{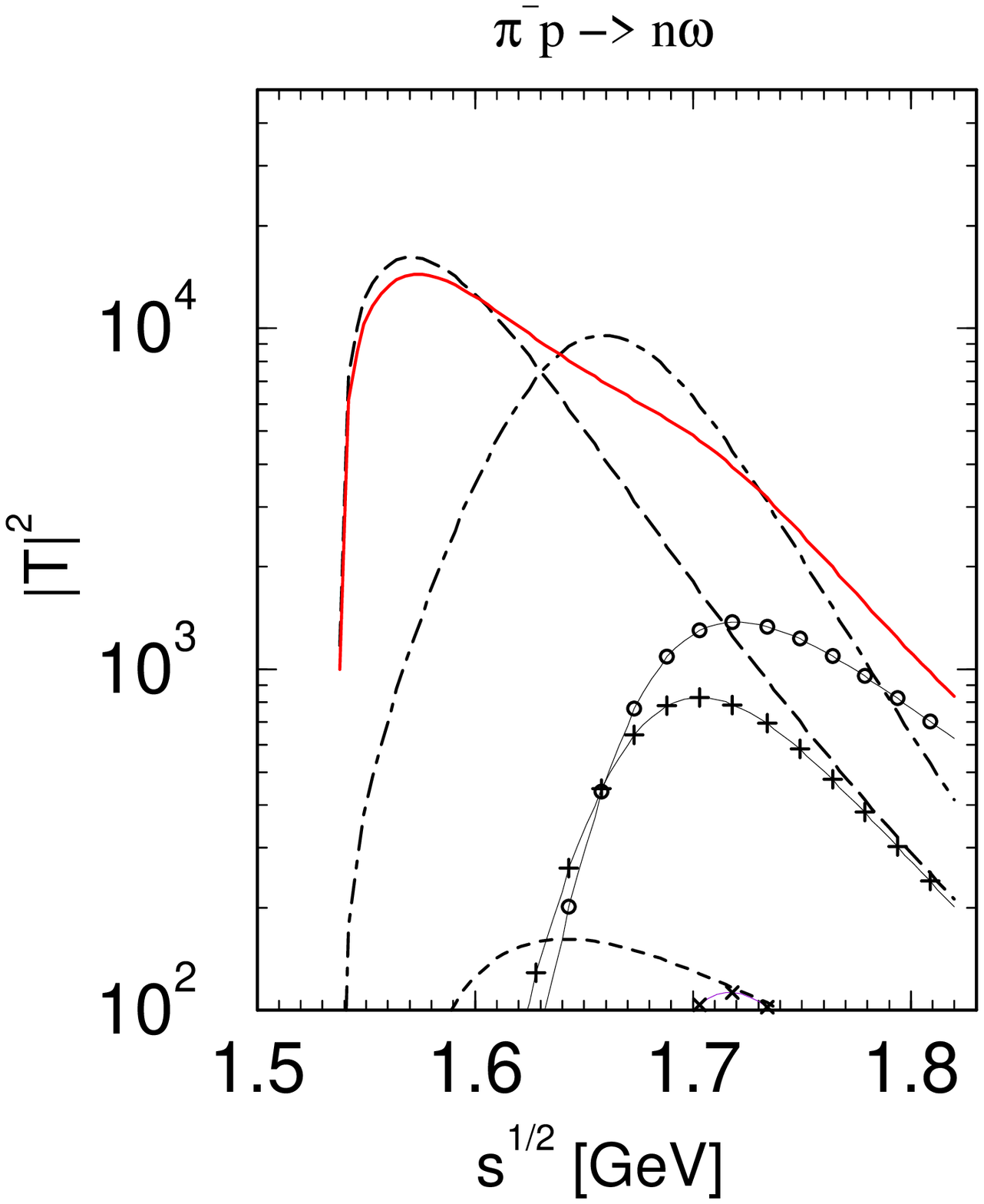}}
~\vskip -12mm
\caption{Invariant matrix elements squared as a function
of the energy ($M_{e^+ e^-} = 0.6$ GeV, $\theta = 30^o$).}
\label{fig4}
\end{figure}

\section{COMBINED ANALYSIS OF $\omega$ AND $\phi$ PRODUCTION:
OZI RULE}

 Along the above outlined approach one can also perform a combined
 study of near-threshold $\omega$ and $\phi$ production. Here, the
 inclusion of the $t$ channel with the $\rho$ meson exchange 
 becomes important.
 We found \cite{we1} that within the conventional meson - nucleon
 dynamics the deviation from the standard OZI rule violation
 in these processes can naturally be explained without invoking any
 strangeness content in the nucleon when including properly the
 baryon resonances. This conclusion is addressed mainly to the
 ratio of the $\phi\rho\pi$ and $\omega\rho\pi$ coupling
 constants, because the resonance contributions to the 
 total cross section of the $\phi$ meson production is rather weak.
 The status of the OZI rule for $\phi NN$ and $\phi NN^*$
 interactions may be checked by considering
 more sensitive (spin dependent) variables and
 in $\omega$ and $\phi$ photoproduction at large angles
\cite{A.Titov}. 

\section{SUMMARY}

An experimental study of the reactions
$\pi^-p\to n e^+e^-$ and $\pi^+n \to p e^+ e^-$
allows to pin down the isoscalar part of the electromagnetic
current and to reveal the role of the subthreshold
$N \omega$ resonances. Details of the $\rho - \omega$ interference
depend on the transition couplings nucleon - baryon resonance - meson.
The knowledge of the mentioned reactions is a necessary
prerequesit for analyzing forthcoming data of the
reaction $\pi A \to X e^+ e^-$ to be studied at HADES.

\end{document}